# Acoustic Transmitters for Underwater Neutrino Telescopes

**Miguel Ardid \*, Juan A. Martínez-Mora, Manuel Bou-Cabo, Giuseppina Larosa, Silvia Adrián-Martínez and Carlos D. Llorens**

Research Institute for Integrated Management of Coastal Areas – IGIC, Universitat Politècnica de València, Paranimf 1, E–46730 Gandia, València, Spain; E-Mails: jmmora@upv.es (J.A.M.-M.); maboca3@doctor.upv.es (M.B.-C.); giula@upv.es (G.L.); siladmar@upv.es (S.A.-M.); cdavid@upv.es (C.D.L.)

\* Author to whom correspondence should be addressed; E-Mail: mardid@fis.upv.es; Tel.: +34-962-849-314; Fax: +34-962-849-309.

**Abstract:** In this paper acoustic transmitters that were developed for use in underwater neutrino telescopes are presented. Firstly, an acoustic transceiver has been developed as part of the acoustic positioning system of neutrino telescopes. These infrastructures are not completely rigid and require a positioning system in order to monitor the position of the optical sensors which move due to sea currents. To guarantee a reliable and versatile system, the transceiver has the requirements of reduced cost, low power consumption, high pressure withstanding (up to 500 bars), high intensity for emission, low intrinsic noise, arbitrary signals for emission and the capacity of acquiring and processing received signals. Secondly, a compact acoustic transmitter array has been developed for the calibration of acoustic neutrino detection systems. The array is able to mimic the signature of ultra-high-energy neutrino interaction in emission directivity and signal shape. The technique of parametric acoustic sources has been used to achieve the proposed aim. The developed compact array has practical features such as easy manageability and operation. The prototype designs and the results of different tests are described. The techniques applied for these two acoustic systems are so powerful and versatile that may be of interest in other marine applications using acoustic transmitters.

## 1. Introduction

In this paper, different R&D studies and prototypes on acoustic transmitters are presented, that were conducted in the context of deep-sea neutrino telescopes. Acoustics are used in this type of facilities mainly in two areas: the acoustic positioning system used to monitor the positions of the optical sensors placed throughout the detector [1], and systems for acoustic neutrino detection technique [2], which is currently under study. Our research group has some responsibilities in these areas in two European partnerships for the design, construction and operation of undersea neutrino telescopes: ANTARES [3] (which is now operational and collecting data), and KM3NeT [4] (which is in the preparatory phase, that is definition and validation of the final design of the facility, and dealing with the legal and financial aspects for the construction). Conceptually both the ANTARES and KM3NeT projects are quite similar. The physics goals of deep-sea neutrino telescopes center on the fields of astronomy, dark matter, cosmic rays and high energy particle physics. Besides, these facilities also hold different equipment for long-term continuous monitoring of environmental parameters interesting in several fields of Earth-Sea science such as biology, oceanography, geology, *etc.*

A deep-sea neutrino telescope is composed of several semi-rigid structures named Detection Units (DUs) that are anchored to the sea bed at great depths (events induced by neutrinos from astrophysical sources must be distinguished from other kinds of events which originate in the Earth's atmosphere). The mechanical structures of the DU contain several Optical Modules (OMs) with photomultiplier sensors detecting the Cherenkov light emitted by muons that are generated in neutrino interactions with matter near or in the detector. Since the DU structure is hundreds of meters high and is held vertically by a buoy located at the top of the structure, underwater sea currents produce inclination of the structures and thus the OMs can be displaced several meters from their nominal positions. For this reason, a positioning system is needed in order to monitor the positions of OMs. In particular, ANTARES is deployed at about 2,500 m depth, about 40 km off the coast of Toulon (France), and has 12 DU (Lines) with a separation between neighbouring lines of about 70 m. Each DU has 25 floors (storeys) with three OMs per storey. The vertical distance between adjacent storeys is 14.5 m. This layout—a 3-dimensional array of OMs over a volume of about 0.05 $km^3$—allows for a precise reconstruction of the muon tracks and thus of the primary neutrinos. KM3NeT will have a volume of several cubic kilometres becoming the next generation of deep-sea neutrino telescopes.

The performance of the telescope (particularly, an accurate reconstruction of the muon track) is highly sensitive to the knowledge of the OM relative positions. Hence it is necessary to monitor the relative positions of all OMs with accuracy better than 20 cm, equivalent to the ~1 ns precision of timing measurements [5]. The muon trajectory reconstruction and determination of its energy also require the knowledge of the OMs orientation with a precision of a few degrees. In addition, a precise absolute positioning of the whole detector has to be guaranteed in order to point to individual neutrino sources in the sky. For all these purposes, a positioning calibration system is needed. This system includes an Acoustic Positioning System (APS) composed of synchronized acoustic transceivers, anchored on fixed known positions at the sea bottom, and receiver hydrophones attached to the DUs structure, close to the position of the OMs. By measuring the time of flight between transceivers and hydrophones, and knowing the sound speed, it is possible to determine the distances between them. The position of the hydrophones is determined by a triangulation method using different transceivers. Using a mechanical model that explains the mechanical behaviour of the DUs and using the position information of the hydrophones (individual points on the DUs), it is possible to reconstruct the position of the OMs with the required accuracy [1]. The first part of this paper (Section 2) summarises the work, studies results and conclusions achieved in last years in the development process of an acoustic transceiver for the APS in the framework of the KM3NeT neutrino telescope. The implementation of the proposed transceiver into the detector is currently evaluated.

The other application of acoustic transmitters presented in this paper is related to the acoustic detection of neutrinos. The possibility of detecting ionizing particles by acoustic techniques was first pointed out by Askarian in 1957. The thermo-acoustic model predicts that an acoustic signal can be produced from the interaction of an Ultra-High-Energy (UHE) neutrino in water. This interaction produces a particle cascade that deposits a high amount of energy in a relatively small volume of the medium, which instantaneously forms a heated volume that gives rise to a measurable pressure signal [2]. Different simulations have been made on the acoustic signal generation and propagation. Details can be found in [6] and references therein. For this work, some reference figures for calibration purposes suffice. On average, 25% of the neutrino energy is deposed by a hadronic shower in a small, almost cylindrical, volume of a few cm in radius and several meters in length. The generated pressure signal has a bipolar shape in time and 'pancake' directivity, this means a flat disk emission pattern perpendicularly to the axis defined by the hadronic shower. As a reference example, we will consider that at 1 km distance, in direction perpendicular to a $10^{20}$ eV hadronic shower, the acoustic pulse has about 0.1 Pa peak-to-peak amplitude and about 40 μs width. With respect to the directivity pattern, the opening angle of the pancake is about 1°.

Both experiments, ANTARES and KM3NeT, consider acoustic detection as a possible and promising technique to cover the detection of UHE neutrinos with energies above $10^{18}$ eV. Also the combination of these two neutrino detection techniques to achieve a hybrid underwater neutrino telescope is possible, especially considering that the optical neutrino techniques need acoustic sensors for positioning purposes. Moreover, ANTARES has an acoustic detection system called AMADEUS that can be considered as a basic prototype to evaluate the feasibility of the neutrino acoustic detection technique. This system is a functional prototype array [7] composed of six acoustic storeys, three of them located on a special DU with instrumentation equipment (Instrumentation Line) and the other three on the 12th DU. Each storey contains six acoustic sensors. The system is operational and taking data. Despite all of the sensors having been calibrated in the laboratory, it would be desirable to have a compact calibrator that may allow for "*in situ*" monitoring of the detection system, to train the system and tune it, in order to improve its performance to test and validate the technique, as well as determining the reliability of the system [8]. The compact transmitter proposed may mimic the signature of a UHE neutrino interaction considering the high directivity of the bipolar acoustic pulse and, in addition, to have small geometrical dimensions that facilitates deployment and operation. In this paper, the studies and prototype developments towards such a transmitter based on the parametric acoustic sources techniques are presented in Section 3.

We believe that these transmitters (with slight modification) may also be used in other applications, such as marine positioning systems, alone or combined with other marine systems, or integrated in different Earth-Sea Observatories, where the localization of the sensors is an issue. Some of these techniques can be also applied for SONAR developments or acoustic communication, especially when very directive beams are required and/or signal processing techniques are needed. In that case, the experience gained from this research can be of great benefit for other applications beyond underwater neutrino telescopes.

## 2. Transceiver Development for the KM3NeT APS

The APS for the future KM3NeT neutrino telescope consists of a series of acoustic transceivers distributed on the sea bottom and receivers located on the DUs near the optical modules. Each of these acoustic transceivers is composed of a transducer and an electronic board to manage it. These two components and the performed tests on them are presented in the following sections.

*2.1. The Acoustic Sensor*

The acoustic sensor has been selected to meet the specifications of the KM3NeT positioning which are: withstanding high pressure, a good receiving sensitivity and transmitting power capability, near omnidirectionality, low electronic noise level, a high reliability, and also affordable pricing for the units needed. Among the different options, we have selected a Free Flooded Ring (FFR) transducer SX30 model (FFR-SX30) manufactured by Sensor Technology Ltd. (Collingwood, Canada, http://www.sensortech.ca). FFR transducers have ring geometrical form maintaining the same hydrostatic pressure inside and outside, whilst reducing the change of the properties of the piezoelectric ceramic under high hydrostatic pressure. For these reasons they are a good solution to the deep submergence problem [9]. FFR-SX30s are efficient transducers that provide reasonable power levels over wide range of frequencies, and deep ocean capability. They work in the 20–40 kHz frequency range with an outer diameter of 4.4 cm, an inner diameter of 2 cm and a height of 2.5 cm. They can be operated in deep-sea scenarios with a transmitting and receiving voltage response at 30 kHz of 133 dB re 1 µPa/V at 1m and −193 dB re 1 V/µPa, respectively. The maximum input power is 300 W with 2% duty cycle. These transducers are simple radiators and have an omnidirectional directivity pattern in the plane perpendicular to the axis of the ring (XY-plane), whilst the aperture angle in the other planes depends on the length of the cylinder (XZ-plane), which is of 60° for the SX30 model. The cable on the free-flooded rings is a 20 AWG type, which is thermoplastic elastomer (TPE) insulated. The cable is affixed directly to the ceramic crystal. The whole assembly is then directly coated with epoxy resin. Both the epoxy resin and the cable are stable in salt water, oils, mild acids and bases. The cables are therefore not water blocked (fluid penetration into the cable may cause irreversible damage to the transducer). For this reason, and following KM3NeT technology standards, the FFR-SX30 hydrophones have been over-moulded with polyurethane material to block water and to facilitate its fixing and integration on mechanical structures. Figure 1 shows pictures of the FFR-SX30 transducer with and without over-moulding.

With the objective to study the possibility of using these transducers as emitters in the APS of the KM3NeT neutrino telescope several studies have been performed. In the following, the results of tests carried out in our laboratory are presented, which characterize the transducers in terms of the transmitting and receiving voltage responses as a function of the frequency and as a function of the angle (directivity pattern). For the tests omnidirectional calibrated transducers, model ITC-1042 (transmitting voltage response 148 dB re 1 µPa/V @ 1m; http://www.itc-transducers.com) and RESON-TC4014 (receiving voltage response −186 dB ± 3dB re 1 V/µPa; http://www.reson.com), were used as a reference emitter and receiver, respectively. The measurements have been performed in a $87.5 \times 113 \times 56.5$ cm$^3$ fresh-water tank using 10-cycle tone burst signals and a distance separation between transducers of about 10 cm. The emitter was fed with moderate voltage (less than 10 V) in order to avoid transient effects, and stable results were obtained with statistical uncertainties below 1 dB.

**Figure 1.** View of the Free Flooded Ring hydrophone: (left) without and (right) with over-moulding.

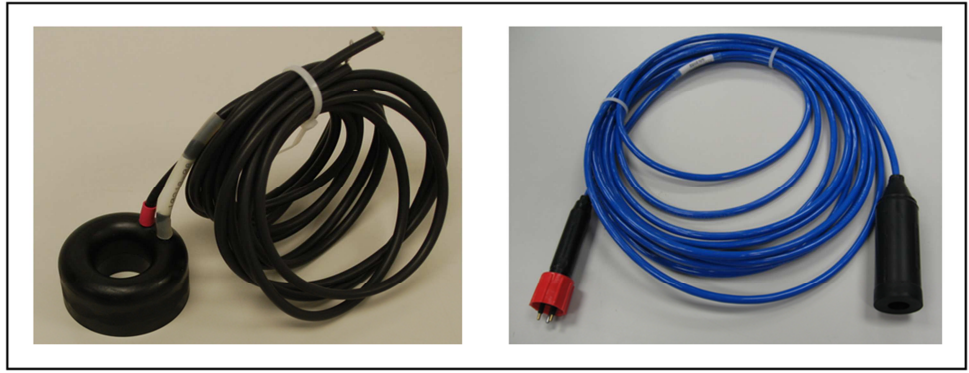

Figure 2 shows the transmitting voltage response and the receiving voltage response of the FFR-SX30 hydrophone as a function of the frequency (measured in the XY-plane, *i.e.*, perpendicular to the axis of the transducer). The results are consistent with the values given by the manufacturer and show that there are no strong irregularities in the 20–40 kHz frequency range.

**Figure 2.** Transmitting and receiving voltage response of the FFR-SX30 hydrophones as a function of the frequency. The uncertainties on the measurements are 1.0 dB.

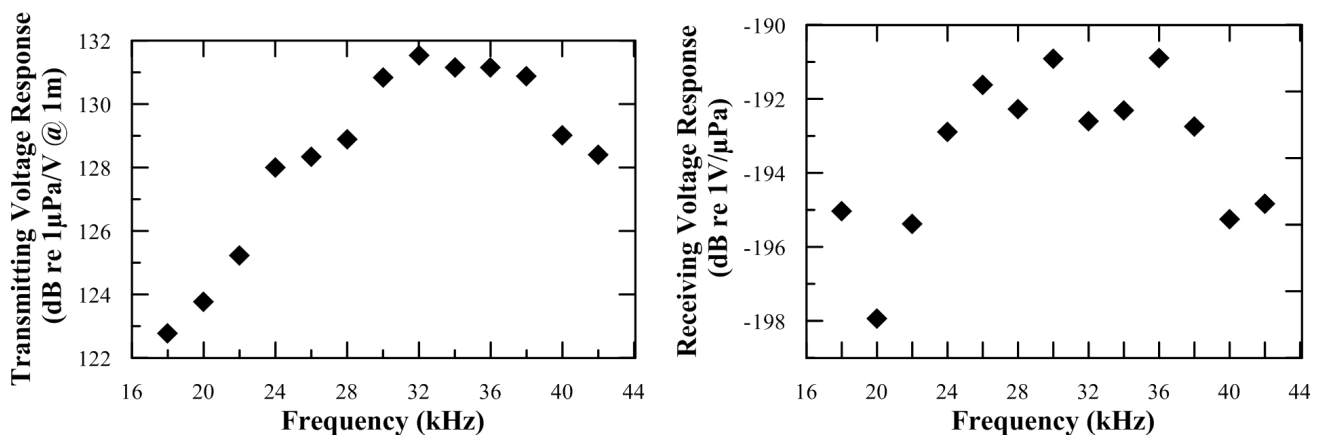

Figure 3 shows the transmitting voltage response and the receiving voltage response of the FFR-SX30 hydrophones as a function of the angle using a 30 kHz tone burst signal (measured in the XZ-plane where 0° corresponds to the direction opposite to cables). As expected, a minimum of sensitivity appears at an angle of ~30°. The variations of sensitivity are about 5 dB (almost 50%), which is a noticeable value, but can be handled without major problems for the KM3NeT APS application.

**Figure 3.** Transmitting and receiving voltage response of the FFR-SX30 hydrophones as a function of the angle in the XZ-plane. The uncertainties on the measurements are 1.0 dB.

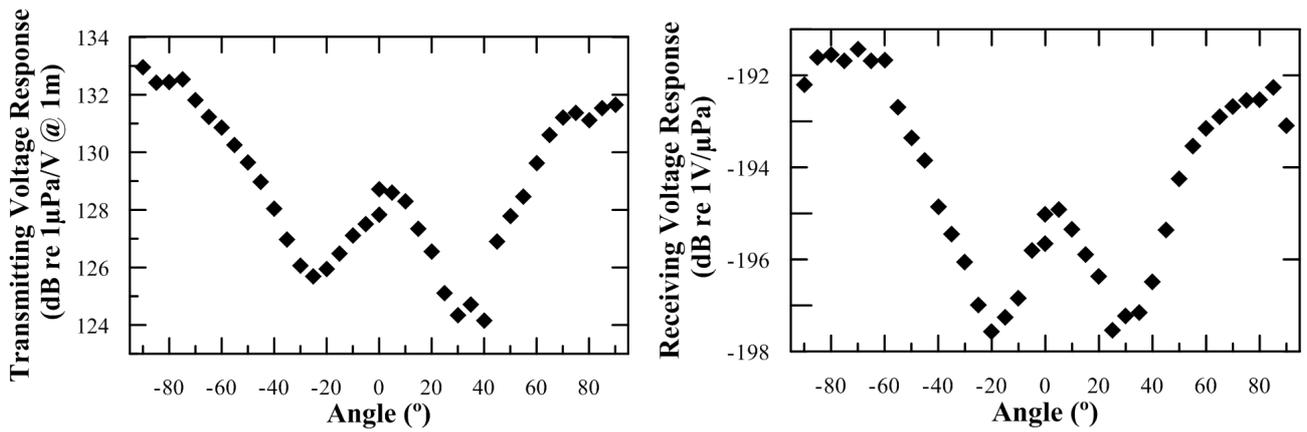

One of the most important aspects that should be validated is the operability of the selected transducers under high pressure. For this reason a measurement campaign using the large hyperbaric tank at the IFREMER research facilities in Plouzanne (near Brest, France) was performed [10]. The behaviour of the FFR-SX30 transducers under different values of pressure were measured, in the frequency range of interest [24 kHz–40 kHz] and the relative acoustic power variations were observed. Figure 4 shows the results obtained from these measurements for the acoustic transmission between two FFR-SX30 transducers. From the measurements we can conclude that these transducers are quite stable with depth. The small variations that were observed are not problematic for the KM3NeT APS application.

**Figure 4.** Pressure dependence of the FFR-SX30 hydrophones as a function of the frequency. The uncertainties of the measurement are 1.0 dB.

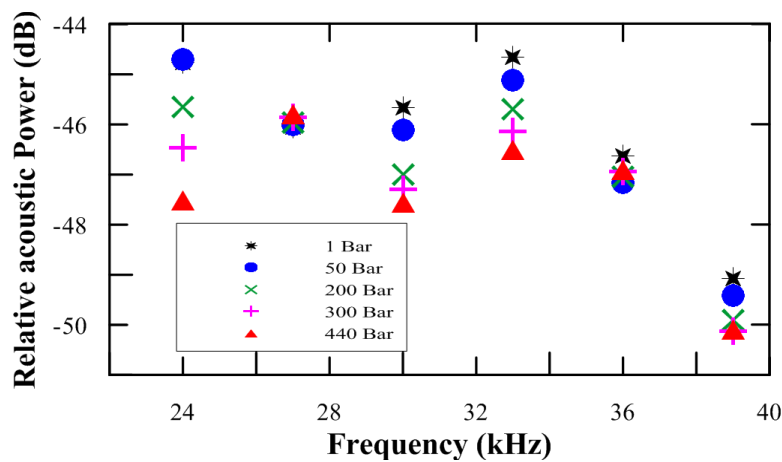

*2.2. The Sound Emission Board*

Dedicated electronics, such as the Sound Emission Board (SEB) [11], have been developed for the communication with and the configuration of the transceiver, and furthermore, to control the emission and reception. Regarding the emission, it is able to generate signals for positioning with enough acoustic power that they can be detected by acoustic receivers at 1−1.5 km away from the emitter. Moreover, it stores electric energy for the emission, and can switch between emission and reception modes. The solution adopted is specially adapted to the FFR-SX30 transducers and is able to feed the transducer with a high amplitude of short signals (a few ms) with arbitrary waveforms. It has the capacity of acquiring the received signal as well. The board prototype diagram is shown in Figure 5. It consists of three parts: the communication and control which contains a micro-controller dsPIC (blue part), the emission part, constituted by the digital amplification plus the transducer impedance matching (red part), and the reception part (green part). In the reception part a relay controlled by the dsPIC switches the mode and feeds the signal from the transducer to the receiving board of the positioning system.

**Figure 5.** View and diagram of the Sound Emission Board.

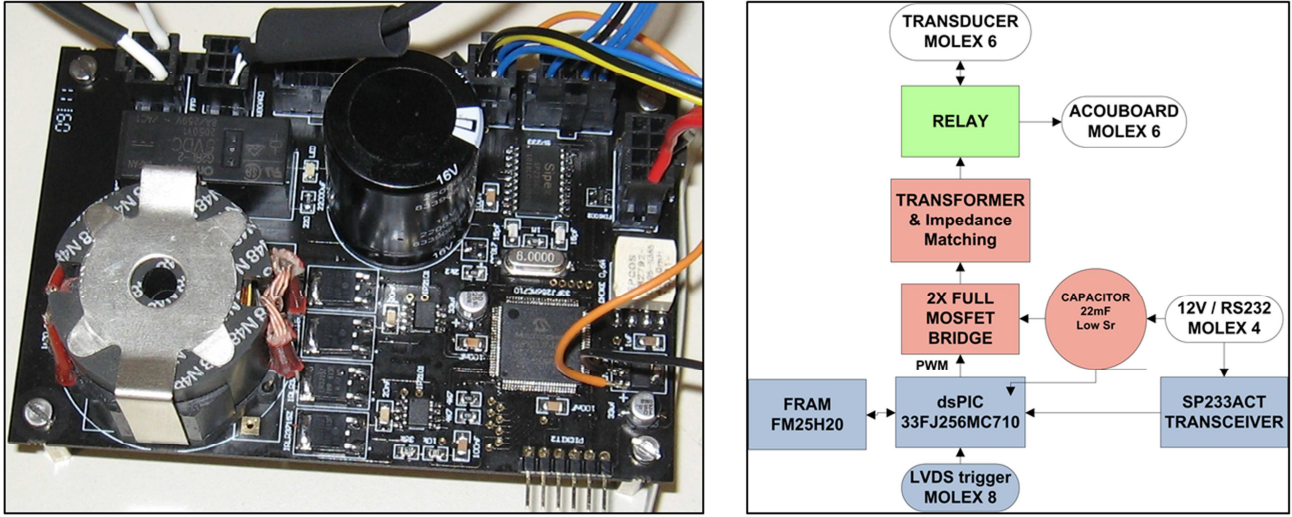

The SEB has been designed for low-power consumption and it is adapted to the neutrino infrastructure using power supplies of 12 V and 5 V with a consumption of 1 mA and 100 mA, respectively, furnished by the power lines of the neutrino telescope infrastructure. With this, a capacitor with a very low equivalent series resistance and 22 mF of capacity is charged storing the energy for the emission. The charge of this capacitor is monitored using the input of the ADC of the micro-controller. Moreover, the output of the micro-controller is connected through a 2× Full MOSFET driver and a MOSFET full bridge; this is successively connected to the transformer with a frequency and duty cycle set through the micro-controller. The transformer is able to increase the voltage of the input signal to an up to 430 Vpp output signal.

The reception part of the board has the additional possibility to directly apply an anti-aliasing filter and return the signal to an ADC of the microcontroller. This functionality may be very interesting not only in the frame of the neutrino telescopes, but also for the implementation in different underwater applications, such as affordable sonar systems or echo-sounders.

The micro-controller runs the program for the emission of the signals and all the parts of the board control. The signal modulation is done using the Pulse-Width Modulation (PWM) technique [12] which permits the emission of arbitrary intense short signals. The PWM carry frequency of the emission signal is 400 kHz, frequencies up to 1.25 MHz have been successfully tested. The basic idea of this technique is to modulate the signal digitally at high frequency using different width of pulses, the lower frequency signal is recovered using a low-pass filter. A full H-Bridge is also used to increase the amplitude of the signal. The communication of the board with the control PC for its configuration is established through the standard RS232 protocol using a SP233 adapter on the board. To provide a very good timing synchronization the emission is triggered using a LVDS signal.

In summary, the board was designed for easy integration in neutrino telescope infrastructures, it can be configured from shore and can emit arbitrary intense short signals, or act as receiver with very good timing precision (the measured latency is 7 μs with a stability better than 1 μs), as shown in KM3NeT APS joint tests [13].

*2.3. Tests of the Transceiver Prototype*

The transceiver has been tested in a fresh water tank in the laboratory, in a pool and in shallow sea water. After characterisation, it has been integrated in the instrumentation line of the ANTARES neutrino telescope for in situ tests in the deep-sea. In the following, the activities and results of these tests are described.

The measurement tests in the laboratory have been performed firstly in a fresh-water tank of $87.5 \times 113 \times 56.5$ cm$^3$, and secondly in a water pool of 6.3 m length, 3.6 m width and 1.5 m depth. We have tested the system using the over-moulded FFR-SX30 hydrophone and the SEB. The complete over-moulding of the transducer has been done by McArtney-EurOceanique SAS. Moreover, a 10 meter-length cable 4021 type (http://www.macartney.com/) has been moulded onto a free issued hydrophones plus one connector type OM2M with its locking sleeves type DLSA-M/F. The mouldings are made of polyurethane, the connector body of neoprene and the locking sleeve of plastic.

Some changes to the SEB board have been made to integrate the system into the ANTARES neutrino telescope and also to test the system *in situ* at 2,475 m depth. For simplicity and due to limitations in the instrumentation line, it was decided to test the transceiver only as an emitter. The receiver functionality will be tested in other in situ KM3NeT tests. The changes made to the SEB are: to eliminate the reception part, to adapt the RS232 connection to RS485 connection and to implement the instructions to select the kind of signals to emit matching the procedures of the ANTARES DAQ system [3].

To test the system, the transceiver has been used with different emission configurations in combination with the omnidirectional transducers ITC-1042 and RESON-TC4014, which were used as emitter and receiver, respectively. Different signals have been used (tone bursts, sine sweeps, maximum length sequence (MLS) signals, *etc.*) to view the performance of the transducer in different situations.

Figure 6 shows the transmitting acoustic power of the transceiver as a function of the frequency (measured in the XY-plane). The transmitting acoustic power of the transceiver as a function of the angle (directivity pattern) using a 30 kHz short tone burst signal is also shown (measured in the XZ-plane, 0° corresponds to the direction opposite to cables). The measurements have been done in similar conditions to those of Figures 2 and 3.

Comparing the Receiving and Transmitting Voltage Response of the FFR-SX30 over-moulded with and without over-moulding, a loss of ~1–2 dB is observed for the over-moulded hydrophone. Figure 6 shows that the result for the transmitting acoustic power in the 20–50 kHz frequency range is in the 164–173 dB re 1 μPa @ 1 m range, in agreement with the electronics design and the specifications needed. Despite this, acoustic transmitting power may be considered low in comparison with the ones used in Long Base Line positioning systems, which usually reach values of 180 dB re 1 μPa @ 1 m. The use of longer signals in combination with a broadband frequency range and signal processing techniques will allow us to increase the signal-to-noise ratio, and therefore allow for the possibility of having an acoustic positioning system of the 1 μs stability (~1.5 mm) order over distances of about 1 km, using less acoustic power, *i.e.*, minimizing the acoustic pollution. In order to study the transceiver over longer distances and also the possibilities of the signal processing techniques, tests were designed in shallow sea water at the Gandia Harbour (Spain). Here, the transceiver was used as an emitter and another FFR-SX30 hydrophone as a receiver with a distance of 140 m between them. The environment was quite hostile for acoustic measurements with a high level of noise existence and multiple reflection sites. However, our analysis showed that the use of broadband signals, MLS and sine sweep signals, is a very useful tool to increase the signal-to-noise ratio and allows for a better distinction of the direct signal from reflections. The latter could be misinterpreted as the direct ones giving a bad detection time [14].

**Figure 6.** Transmitting acoustic power of the transceiver as functions of frequency and angle, respectively. The uncertainties on the measurements are 1.0 dB.

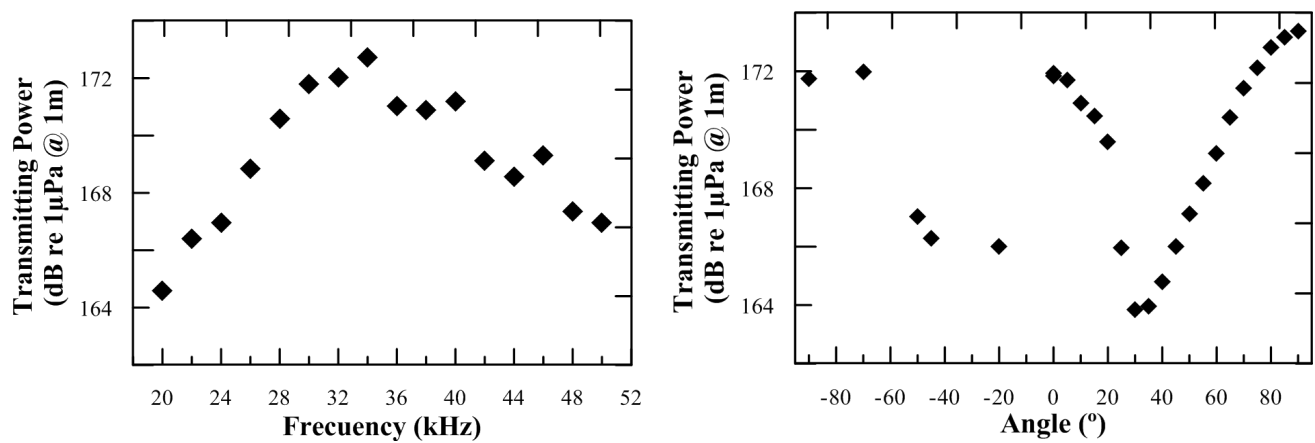

The system has been finally integrated into the active anchor of the Instrumentation Line of ANTARES. In particular, the SEB was inserted in a titanium container containing a laser and other electronic boards used for timing calibration purposes. A new functionality for the microcontroller was implemented to control the laser emission as well. The FFR-SX30 hydrophone was fixed to the base of the line at 50 cm from the standard transmitter of the ANTARES positioning system [1] with the free area of the hydrophone looking upwards.

Figure 7 shows some pictures of the final integration of the system in the anchor of the Instrumentation Line of ANTARES. Finally, the Instrumentation Line was successfully deployed at 2,475 m depth on 7th June 2011 at the nominal target position. The connection of the Line to the Junction Box, will be made when the ROV (Remotely Operated Vehicle) will be available (probably in April 2012). Afterwards the transceiver will be tested in real conditions.

**Figure 7.** Picture of the anchor of the Instrumentation Line of ANTARES showing the final integration of the transceiver. Details of the FFR-SX30 hydrophone with its support and of the titanium container housing the SEB are also shown.

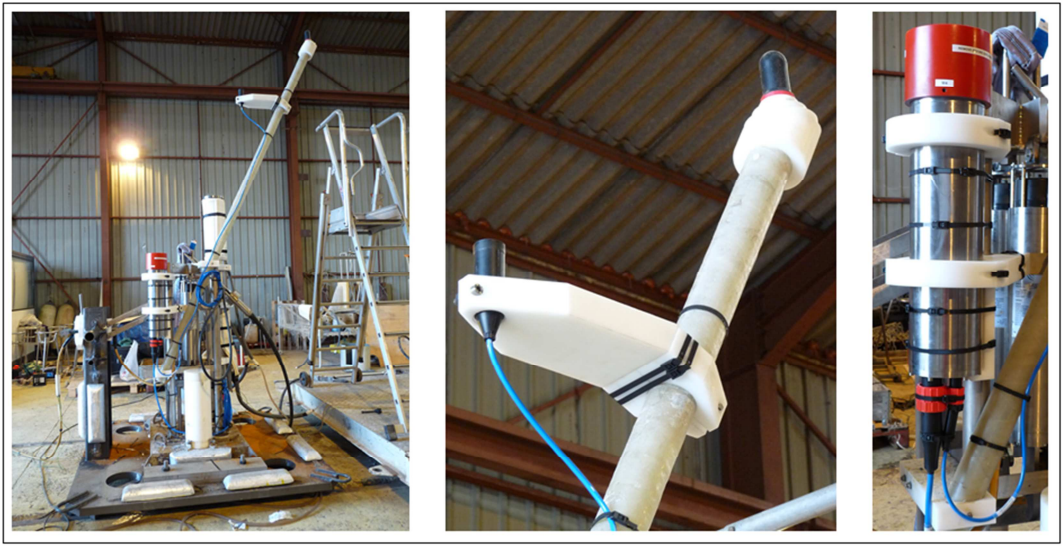

## 3. Compact Transmitter for Acoustic UHE Neutrino Detection Calibration

In this section we present the R&D studies based on parametric acoustic sources techniques in order to develop a compact transmitter prototype for the calibration of acoustic neutrino detection arrays. The aim is to have a very versatile calibrator that not only is able to generate neutrino-like signals, but is able as well to calibrate the sensitivity of the acoustic sensors of the telescope, to check and train the feasibility of the UHE neutrino acoustic detection technique and to generate signals either for positioning or monitoring environmental parameters acoustically [8]. Moreover, a compact solution for the calibrator will result, most probably, in a system easier to install and deploy in undersea neutrino telescopes.

*3.1. Parametric Acoustic Sources*

Acoustic parametric generation is a well-known non-linear effect that was first studied by Westervelt [15] in the 1960s. This technique has been studied quite extensively since then, being implemented in many applications in underwater acoustics, specifically to obtain very directive acoustic sources. The acoustic parametric effect occurs when two intense monochromatic beams with two close frequencies travel together through the medium (water, for example). Under these conditions in the regime of non-linear interaction, secondary harmonics appears with the sum, difference, and double spectral components. One application of this effect is to obtain low-frequency very directive beams from very directive high-frequency beams, as secondary parametric beams have similar directivity pattern as the primary beams. Directive high-frequency beams are much easier to obtain, as the needed dimensions of the emitter scale with the wavelength. The technique application for a compact calibrator presents two difficulties: on one hand, it is a transient signal with broad frequency content, on the other hand, the directivity has cylindrical symmetry. To deal with transient signals it is possible to generate a signal with 'special modulation' at a larger frequency in such a way that the pulse interacts with itself while travelling along the medium, providing the desired signal. In our particular case the desired signal would be a signal with bipolar shape signal in time. Theoretical and experimental studies of parametric generation show that the shape of the secondary signal follows the second time derivative in time of envelope of the primary signal [16], following the equation:

$$p(x,t) = \left(1 + \frac{B}{2A}\right) \frac{P^2 S}{16\pi\rho c^4 \alpha x} \frac{\partial^2}{\partial t^2} \left[ f\left(t - \frac{x}{c}\right) \right]^2 \quad (1)$$

where *P* is the pressure amplitude of the primary beam pulse, *S* the surface area of the transducer, *f(t-x/c)* is the envelope function of the signal, which is modulated at the primary beam frequency, *x* is the distance, *t* is the time, *B/A* the non-linear parameter of the medium, *ρ* is the mass density, *c* the sound speed and *α* is the absorption coefficient.

Parametric acoustic sources have some properties that are very interesting to be exploited in our acoustic compact calibrator:

- It is possible to obtain narrow directional patterns at small overall dimensions of primary transducer.
- The absence of side lobes in a directional pattern of the difference frequency.
- Broad band of operating frequencies of radiated signals.
- Since the signal has to travel long distances, primary high-frequency signal will be absorbed.

*3.2. Evaluation of the Technique for the Application Proposed*

(a) Planar transducers

The first study to evaluate the parametric acoustic sources technique for the emission of acoustic neutrino like signal was to try to reproduce the bipolar shape of the signal. This study was done using planar transducers and was described in [17]. The results in terms of the bipolar signal generation from the primary beam signal, the studies of the signal shape, the directivity patterns obtained, the evidence of the secondary non-linear beam generated in the medium and the checking of the non-linear behaviour with the amplitude of the primary beam have been all coherent with the expectation from theory and demonstrated that the technique could be useful for the development of the compact acoustic calibrator able to mimic the signature of the UHE neutrino interaction. Figure 8 shows the comparison between the emitted signal, the received signal and the secondary signal (obtained using a band-pass FIR filter: with corner frequencies 5 kHz and 100 kHz). The nonlinear behaviour of the secondary beam (filtered signal) in comparison to the primary beam (received signal) is also shown.

(b) Cylindrical symmetry

Once we have been able to reproduce the shape of the desired signal using the parametric technique, the next step was a more detailed study of the 'modulated signal' influence on the secondary beam generated, and on the other hand, to reproduce the 'pancake' pattern of emission desired using a single cylindrical transducer, a Free Flooded Ring SX83 (FFR-SX83) manufactured by Sensor Technology Ltd. (Collingwood, Canada). It has a diameter of 11.5 cm and 5 cm height. This transducer usually works around 10 kHz (the main resonance peak is at 10 kHz), but for our application we use a second peak resonance at about 400 kHz, which is the frequency used for the primary beam. This work is described in [18] and the results obtained agree with the previous ones using planar transducers, but now dealing with the complexity of the cylindrical symmetry generation obtaining a 'pancake' directivity of a few degrees.

**Figure 8.** (**a**) Emitted and received signals. (**b**) Amplitude of the primary and secondary signals as a function of the amplitude of the input signal in the transducer. Statistical uncertainties are very small.

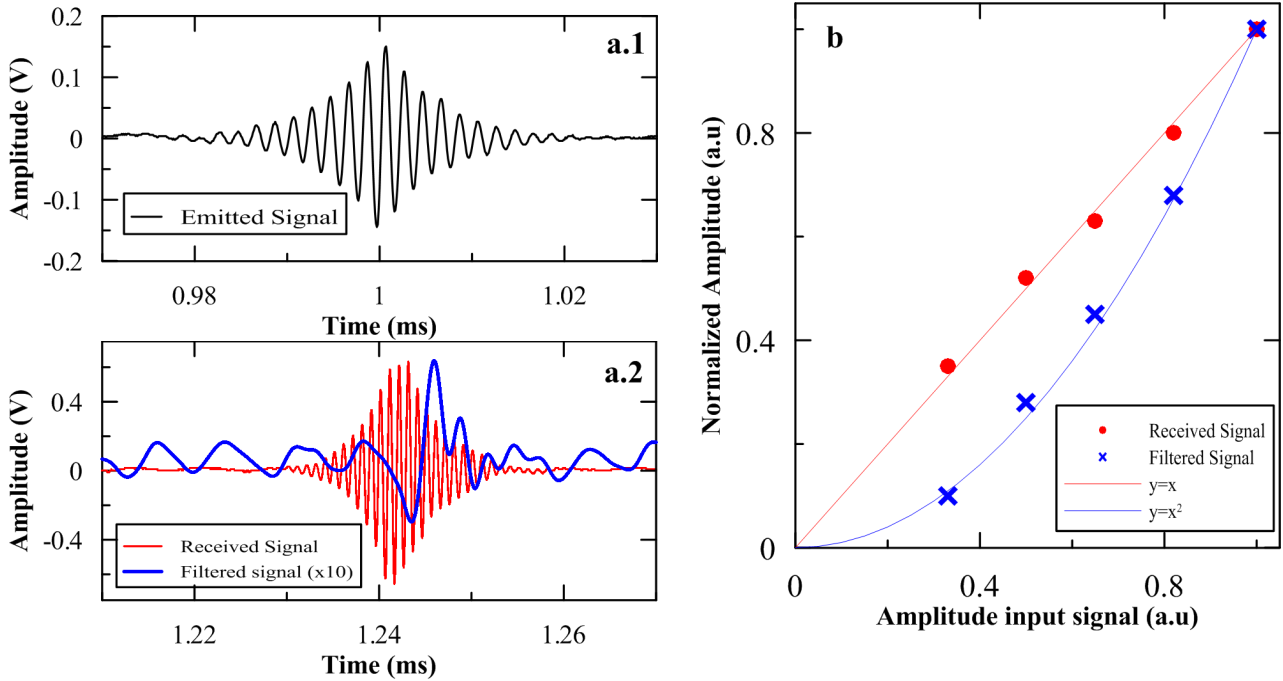

To verify the previous studies over longer distances, new measurements have been made in an emitter-receiver configuration in a pool of 6.3 m length, 3.6 m width and 1.5 m depth using as emitter a single FFR-SX83 transducer. It was positioned at 70 cm depth and the receiver hydrophone used to measure the acoustic waveforms was a spherical omnidirectional transducer (model ITC-1042) connected to a 20 dB gain preamplifier (Reson CCA 1000). With this configuration the receiver presents an almost flat frequency response below 100 kHz with a sensitivity of about −180 dB re 1 V/µPa, whereas it is 38 dB less sensitive at 400 kHz. The larger sensitivity at lower frequencies is very helpful for a better observation of the secondary beam. For these tests, the emitter and receiver are aligned and positioned manually with cm accuracy, which is enough for our purposes. A DAQ system is used for emission and reception. To drive the emission, an arbitrary 14 bits waveform generator (National Instruments, PCI-5412) has been used with a sampling frequency of 10 MHz. This feeds a linear RF amplifier (ENI 1040L, 400W, +55 dB, Rochester, NY, USA) used to amplify the emitted signal. The received signal was recorded with an 8 bit digitizer (National Instruments, PCI-5102) has been used with a sampling frequency of 20 MHz. Later, the recorded data are processed and different band-pass filters are applied to extract the primary and secondary beam signals and the relevant parameters.

Figure 9 summarizes the results of this study by comparing the amplitude of the primary and secondary beams. A different behaviour is observed in the evolution of both beams with the distance. The attenuation exponent for the primary beam is 0.81, which seems a reasonable value considering that we expect an exponent between 0.5 (pure cylindrical propagation) and 1.0 (spherical propagation), being our case an intermediate situation. The attenuation exponent for the secondary beam is 0.50, which also seems reasonable since we expect an attenuation factor significantly lower than that of the primary beam, but higher than half of that exponent. This result is a clear evidence of the secondary bipolar pulse generation by the parametric acoustic sources. The directivity patterns were measured at a distance of 2.3 m. Despite the frequency components are very different both look quite similar (the opening angle differs about 10%).

**Figure 9.** Amplitude of the primary and secondary signals as a function of the distance, and directivity pattern for both beams. Normalization values (**a**) Primary beam: 166 kPa, Secondary beam: 200 Pa; (**b**) Primary beam: 27 kPa, Secondary beam: 80 Pa.

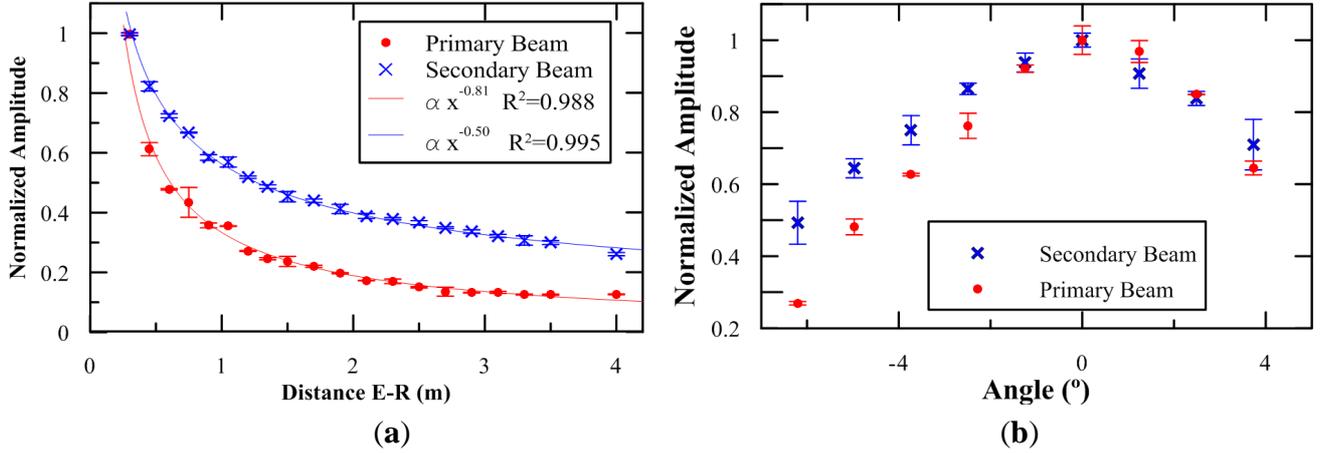

The positive results of the studies with a single transducer confirm that the parametric acoustic sources technique may be applied for the development of a transmitter able to mimic the acoustic signature of a UHE-neutrino interaction. Moreover, the technique presents advantages with respect to other classical solutions: such as the use of a higher frequency in a linear phased array implies that fewer elements are needed on a shorter length scale having a more compact design, and thus, probably easier to install and deploy in undersea neutrino telescopes. A possible drawback of the system is that the parametric generation is not very efficient energetically, but since bipolar acoustic pulses from UHE-neutrino interactions are weak, they can be emulated with reasonable power levels of the primary beams. At this point, it is necessary to design a fully functional array for integration in undersea neutrino telescopes or for application in calibration sea campaigns. There are two aspects that need to be dealt and solved for this: the mechanical design of the array and the necessary electronics to manage the array of acoustic sensors. In the following section, we describe some ideas and work on these aspects.

*3.3. Design of the Compact Array*

Once the single cylindrical transducer had been characterized for the generation of the bipolar pulse using parametric generation in the previous studies, all the required inputs are available for the design of the array able to generate the neutrino-like signal with the 'pancake' directivity with an opening angle of about 1°. Figure 10(a) shows an example of the results for calculations performed by summing the contributions of the different sensors for far distances at different angles. In this example, a linear array of 3 FFR-SX83 transducers with a 20 cm separation from each other is enough to obtain an opening angle of about 1°. Moreover, to estimate the effect of the propagation caused in the bipolar parametric signal, received signals of the experimental measurements of the previous section (a single transducer in the pool), have been extrapolated to 1 km distance. For this an algorithm that works at frequency domain and propagates each spectral component considering the geometric spread of the pressure beams as 1/r and its absorption coefficient [19,20] has been used. The propagation has been done for the sea conditions of the ANTARES site, the value of the parameters are presented in Table 1. Figure 10(b) shows the results of propagating to 1 km the received signal of Figure 9 measured at 2.3 m distance and 0 °. In this case, no filter is applied, the propagation medium acts as a natural filter. High frequencies of the primary beam are absorbed and, at km range, only low frequencies remain. To be exact, there is still a small high-frequency component which is not observed at distances of 1.2 km (or higher). Notice that the high-frequency signal was three orders of magnitude higher than the secondary beam at a 1 m distance. It appears as well a kind of DC offset, it is due to the very low-frequency components of the signal (probably 50 Hz) which are also propagated.

**Figure 10.** (**a**) Pressure signal obtained at different angles for a three-element array. (**b**) Signal obtained by the propagation of the measured signal to a 1 km distance.

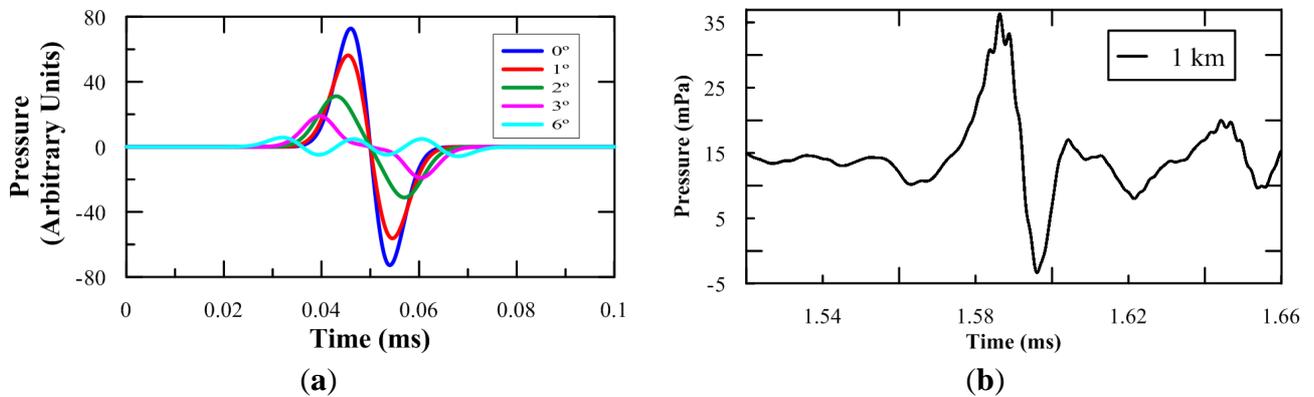

**Table 1.** Parameters used for the propagation and absorption coefficient examples [19,20].

| Depth (m) | Sound speed (m/s) | Salinity (°/₀₀) | Temperature (°C) | pH | Absorption coeff., 25 kHz (Np/m) | Absorption coeff., 400 kHz (Np/m) |
|---|---|---|---|---|---|---|
| 2,200 | 1,541.7 | 38.5 | 13.2 | 8.15 | 0.00042 | 0.00983 |

To conclude this study, for a single element, it is expected to have a bipolar pulse with a 35 mPa amplitude peak-to-peak at 1 km. Considering the array configuration with three elements feeding in to phase with the maximum power it is expected to have, at least, a 0.1 Pa amplitude peak-to-peak, which is a good pressure reference for calibration of neutrino interactions of the $10^{20}$ eV energy range. Therefore, with the goal to reproduce the 'pancake' directivity, to cover long distances and to improve the level of signal of non-linear beam generated at the medium, an array of three elements configuration has been proposed as possible solution. It is composed by 3 FFR-SX83 transducers with a distance between elements of 2 cm, having the active part of the array at a total height of 20 cm. The three elements are maintained in a linear array configuration by using three bars with mechanical holders as shown in Figure 11. The bars can help to hold the array and also to help to orientate it.

**Figure 11.** Picture of the array used for the tests and of the pool during the data taking. The emitter array and the receiver can be observed.

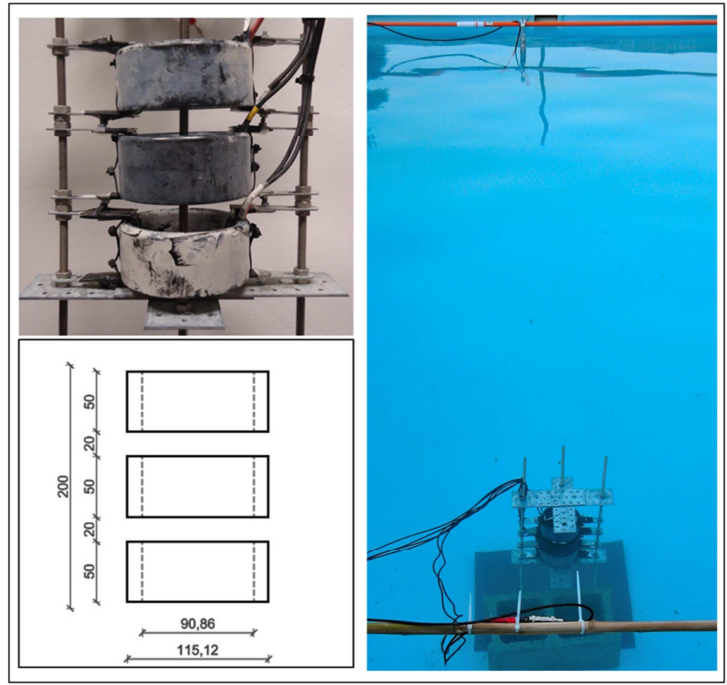

The measurements for the array characterization have been made using the same configuration than for the previous measurements. Figure 11 shows a drawing with the dimensions of the array, and pictures of the array and of the pool during the tests.

The results obtained from the array tests are summarized in Figure 12. An example of a received signal and the primary and secondary beams obtained after applying a band-pass FIR filter filters are shown in Figure 12(a), the secondary beam has been amplified by a factor of 3 for a better visibility. The corner frequencies of the primary beam are 350 kHz and 450 kHz, whereas the cut frequencies for the secondary beam are 5 kHz and 100 kHz. It is possible to see how the reproduction of the signal shape is achieved agreeing the results with our expectations from theory and previous observations. The directivity pattern measured at a longitudinal line defined by the axis of the array is shown in Figure 12(b). These measurements have been made at a 2.7 m distance between the array and the receiver. Notice that the absolute pressure values of Figure 12(b) are smaller than those of Figure 9(b). The reason for this is that the amplifier used is not very efficient to feed the three elements together due to a mismatch of electrical impedances, and therefore each transducer provides a lower pressure beam. Thus, for the final array system, it is very important to work on electronics and have a very good amplifier for each transducer. The design of the electronics for the array is further discussed in next section. In spite of these limitations, Figure 12(b) is quite interesting because it allows studying the angle distribution of both beams. Instead of having for the primary beam a thin peak, a wide peak with a no clear maximum is observed. This is due to the fact that the measurements were done at a distance which cannot be considered very large, and therefore the signals from the different transducers are not totally synchronous at 0°. However, the FWHM measured with the array for the secondary beam is about 7° ($\sigma = 3°$), that is, smaller than for the primary beam, and sensitively smaller than for a single element where the FWHM was about 14° ($\sigma = 6°$). This agrees with the expectations of the kind of calculations described for Figure 10(a), and considering that the signals for the three elements will be better synchronized at far distances (larger than 100 m), a more 'pancake'-like directive pattern with $\sigma \sim 1°$ is expected for far distances.

**Figure 12.** (**a**) Example of a received signal and the primary and secondary beams obtained after applying the band-pass filters (the secondary beam has been amplified by a factor 3 for a better visibility). (**b**) Directivity patterns of primary and secondary beam measured with the array. Normalization values, Primary beam: 5.4 kPa, Secondary beam: 11.5 Pa.

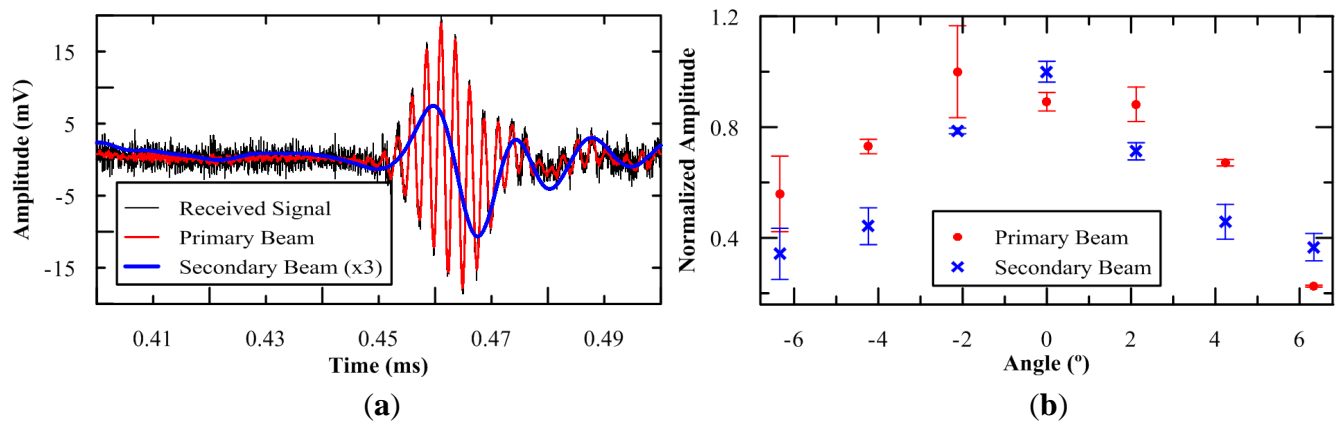

(**a**)  (**b**)

*3.4. Prototype of a Versatile Compact Array*

Our final goal is to have an autonomous and optimized compact system able to carry out several tasks related to acoustics in an underwater neutrino telescope, these tasks being: signal emulation for acoustic neutrino detection arrays, the calibration of acoustic receivers, and even performing positioning tasks, with all of them using the same transmitter. This could reduce the cost and facilitates the deployment in the deep sea. The new developments are orientated towards a mechanical array design improving the directivity and the operation, and the associated electronics to achieve a more powerful autonomous system.

For the prototype, the transducers have been fixed around an axis on flexible polyurethane. This offers water resistance and electrical insulation for high frequency and high voltage applications due to the nature of the cured polymer. Figure 13 shows the compact array prototype. Its compact design is remarkable with an active surface length of 17.5 cm. In order to use it in a future sea campaign with a vessel, a mechanical structure has been built that allows the device to be affixed to a boat, allowing control of the rotation angle. Due to the high directivity of the bipolar pulse, this is a very important point in order to be able to orientate the signal to the direction of the receivers.

**Figure 13.** Compact array prototype and mechanical structure to hold and operate it from a boat.

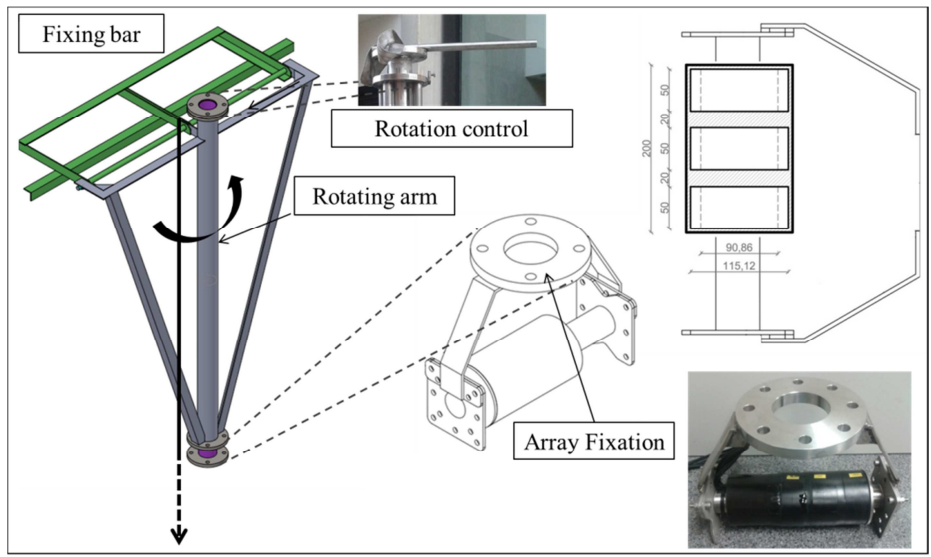

Developments in electronics have been made to achieve an autonomous and optimized compact system able to work in a different frequency ranges and application modes. It has been necessary to develop an electronic device that controls the transmitter which generates and amplifies the signals, in order to have enough acoustic power in the nonlinear regime (parametric generation) and that standard signals can be detected at distant acoustic receivers (for calibration and/or positioning purposes). For this, an electronic board has been developed, based on the same principles as the SEB (described in Section 2.2), that was adapted to the particularities of the transmitter and applications considered. The board is able to communicate, configure the transmitter and control the emission mode (either for high or low frequency operation mode). Similarly to the electronics in the acoustic transceivers for positioning systems [11,21], the PWM technique has been used for the emission of arbitrary intense short signals to emit the necessary 'modulated signal' with the goal to obtain a secondary beam with the specifications desired. Some of the advantages that this technique offers are:

- The system efficiency is improved because the system uses a class D amplification, this means that the transistors are working on switching mode, suffering less power dissipation in terms of heat, and therefore offering a superior performance.
- Simplicity of design. Analogic-digital converters are not needed. It is possible to feed directly the amplifier with digital signal modulated by the PWM technique.
- It is not necessary to install large heat sinks at amplifier transistors, reducing the weight and volume of the electronics system.
- In waiting mode, the power amplifier has a minimum power at idle state that allows storing the energy for the next emission in the capacitor very fast and efficiently.

The diagram of the board prototype is shown in Figure 14. It consists of three parts: the communication and control which contains the micro-controller (in blue) dsPIC33FJ256MC0710 implementing the PWM under motor control technology, the emission constituted by the digital amplification plus the transducer impedance matching (in red), differentiated for each frequency range, and the commuter between the two operation modes (in green). There are three identical parallel boards to manage the three transducers of the compact array. They can be used individually or, in parallel (standard operation mode).

**Figure 14.** Electronics block diagram for the compact array transmitter.

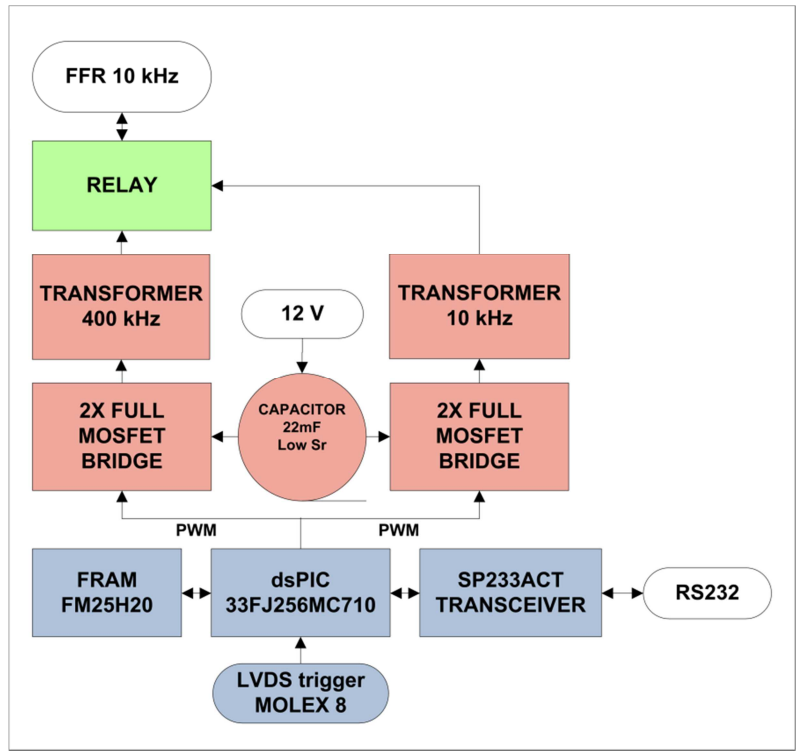

Summarizing, the sound emission board has been designed for an easy integration into neutrino telescope infrastructures, using PWM to emit arbitrary intense short signals, to mimic the acoustic signature of neutrino using parametric acoustic generation, and tone bursts or arbitrary signals with low spectral content for positioning or calibration tasks.

## 4. Conclusions and Future Steps

We have discussed the use and applications of acoustic sensors in underwater neutrino telescopes, and presented the acoustic transmitters developed either for the positioning system of KM3NeT (transceiver of the APS) or for calibration in acoustic neutrino detection systems (compact array prototype).

With respect to the transceiver for the APS, we have shown the results of the tests and measurements performed with FFR-SX30 hydrophones and a custom sound emission board, concluding that the transceiver proposed can be a good solution with the requirements and accuracy needed for such a positioning system.

With respect to the compact array, after showing the R&D studies made, we can conclude that the solution proposed based on parametric acoustic sources could be considered a good tool to generate the acoustic neutrino-like signals, achieving the reproduction of both specific characteristics of the signal predicted by theory: bipolar shape in time and 'pancake directivity'. Moreover, due to the versatility of the transceiver system, this prototype could be implemented to carry out several calibration tasks related to acoustic emission in underwater neutrino telescopes.

Furthermore, we consider that the techniques used for these transmitters are so powerful and versatile that it may be used in other kind of applications, such as marine positioning systems, alone or combined with other marine systems, or integrated in different Earth-Sea Observatories, where the localization of the sensors is an issue. Other applications, such as acoustic communication or SONAR, may benefit from the developments in obtaining very directive sources and in the implementation of signal processing techniques. Moreover, the developments and results may be of great interest for systems with cylindrical acoustic propagation or systems that can work in two operation modes (standard one and parametric acoustic sources one). In that sense, the experience gained from this research can be of interest to open new possible application uses in these areas.

The future work will consist of completing the characterization of the prototype systems, and integrating the transmitters into underwater neutrino telescopes using the framework of the ANTARES and KM3NeT neutrino telescopes. A very useful test for the *in situ* demonstration of the utility of the transmitters is to use them together with the ANTARES-AMADEUS acoustic system, and it is foreseen that these tests will be carried out during 2012.

**Acknowledgments**

This work has been supported by the Ministerio de Ciencia e Innovación (Spanish Government), project references FPA2009-13983-C02-02, ACI2009-1067, AIC10-D-00583, Consolider-Ingenio Multidark (CSD2009-00064). Authors Manuel Bou-Cabo and Silvia Adrián-Martínez thank Multidark for the fellowships. The work has also been funded by Generalitat Valenciana, Prometeo/2009/26, and the European 7th Framework Programme, Grant No. 212525.

**References and Notes**

1. Ardid, M. Positioning system of the ANTARES neutrino telescope. *Nucl. Instr. Meth. A* **2009**, *602*, 174-176.
2. Askariyan, G.A.; Dolgoshein, B.A.; Kalinovsky, A.N.; Mokhov, N.V. Acoustic detection of high energy particle showers in water. *Nucl. Instr. Meth.* **1979**, *164*, 267-278.
3. Ageron, M.; Aguilar, J.A.; Al Samarai, I.; Albert, A.; Ameli, F.; André, M.; Anghinolfi, M.; Anton, G.; Anvar, S.; Ardid, M.; et al. (ANTARES Collaboration). ANTARES: The first undersea neutrino telescope. *Nucl. Instr. Meth. A* **2011**, *656*, 11-38.
4. The KM3NeT Collaboration. *KM3NeT Technical Design Report*; 2010. ISBN 978-90-6488-033-9. Available online: www.km3net.org (accessed on 20 March 2012).
5. Aguilar, J.A.; Al Samarai, I.; Albert, A.; André, M.; Anghinolfi, M.; Anton, G.; Anvar, S.; Ardid, M.; Assis Jesus, A.C.; Astraatmadja, T.; et al. (ANTARES Collaboration). Time calibration of the ANTARES neutrino telescope. *Astrop. Phys.* **2011**, *34*, 539-549.
6. Bevan, S.; Brown, A.; Danaher, S.; Perkin, J.; Rhodes, C.; Sloan, T.; Thompson, L.; Veledar, O.; Waters, D. (ACORNE Collaboration). Study of the acoustic signature of UHE neutrino interactions in water and ice. *Nucl. Instr. and Meth. A* **2009**, *607*, 398-411.
7. Aguilar, J.A.; Al Samarai, I.; Albert, A.; M.; Anghinolfi, M.; Anton, G.; Anvar, S.; Ardid, M.; Assis Jesus, A.C.; Astraatmadja, T.; Aubert, J.-J.; et al. (ANTARES Collaboration). AMADEUS-The acoustic neutrino detection test system of the ANTARES deep-sea neutrino telescope. *Nucl. Instr. Meth. A* **2011**, *626*, 128-143.
8. Ardid, M. Calibration in acoustic detection of neutrinos. *Nucl. Instr. Meth. A* **2009**, *604*, S203-S207.

9. Sherman, C.H.; Butler, J.L. *Transducers ad Array for Underwater Sound*; Springer: New York, USA, 2007.
10. Ardid, M; Bou-Cabo, M.; Camarena, F.; Espinosa, V.; Larosa, G.; Llorens, C.D.; Martínez-Mora, J.A. A prototype for the acoustic triangulation system of the KM3NeT deep sea neutrino telescope. *Nucl. Instr. Meth. A* **2010**, *617*, 459-461.
11. Llorens, C.D.; Ardid, M.; Sogorb, T.; Bou-Cabo, M.; Martínez-Mora, J.A.; Larosa, G.; Adrián-Martínez, S. The Sound Emission Board of the KM3NeT Acoustic Positioning System. *J. Instrum.* **2012**, *7*, C01001.
12. Barr, M. Introduction to Pulse Width Modulation. *Embed. Syst. Program.* **2001**, *14*, 103-104.
13. Simeone, F.; Ameli, F.; Ardid, M.; Bertin, V.; Bonori, M.; Bou-Cabo, M.; Calì, C.; D'Amico, A.; Giovanetti, G.; Imbesi, M.; et al. Design and first tests of an acoustic positioning and detection system for KM3NeT. *Nucl. Instr. Meth. A* **2012**, *662*, S246-S248.
14. Larosa, G; Ardid, M.; Llorens, C.D.; Bou-Cabo, M.; Martínez-Mora, J.A.; Adrián-Martínez, S. Development of an acoustic transceiver for the KM3NeT positioning system. *Nucl. Instr. Meth. A* **2012**, accepted.
15. Westervelt, P.J. Parametric acoustic array. *J. Acoust. Soc. Am.* **1963**, *35*, 535-537.
16. Moffett, M.B.; Mello, P. Parametric acoustic sources of transient signals. *J. Acoust. Soc. Am.* **1979**, *66*, 1182-1187.
17. Ardid, M; Bou-Cabo, M.; Camarena, F.; Espinosa, V.; Larosa, G.; Martínez-Mora, J.A.; Ferri, M. Use of parametric acoustic sources to generate neutrino-like signals. *Nucl. Instr. Meth. A* **2009**, *604*, S208-S211.
18. Ardid, M; Adrián, S.; Bou-Cabo, M.; Larosa, G.; Martínez-Mora, J.A.; Espinosa, V.; Camarena, F.; Ferri, M. R&D studies for the development of a compact transmitter able to mimic the acoustic signature of a UHE neutrino interaction. *Nucl. Instr. Meth. A* **2012**, *662*, S206-S209.
19. Francois, R.E.; Garrison G.R. Sound absorption based on ocean measurements. Part I: Pure water and magnesium sulfate contributions. *J. Acoust. Soc. Am.* **1982**, *72*, 896-907.
20. Francois, R.E.; Garrison G.R. Sound absorption based on ocean measurements. Part II: Boric acid contribution and equation for total absorption. *J. Acoust. Soc. Am.* **1982**, *72*, 1879-1890.
21. Ardid, M; Bou-Cabo, M.; Camarena, F.; Espinosa, V.; Larosa, G.; Llorens, C.D.; Martínez-Mora, J.A. R&D towards the acoustic positioning system of KM3NeT. *Nucl. Instr. Meth. A* **2011**, *626-627*, S214-S216.